\documentclass[aps,twocolumn,showpacs,lengthcheck,preprintnumbers]{revtex4-1} 
\usepackage{amsfonts}
\usepackage{amsmath}
\usepackage{amssymb}
\usepackage{graphicx}
\usepackage{fontenc}
\usepackage{color}

\usepackage[unicode]{hyperref}
\usepackage{microtype}
\hypersetup{
	pdftitle={chiral currents},
	pdfauthor={},
	pdfsubject={},
	colorlinks=true,
	linkcolor=blue,
	citecolor=blue,
	filecolor=black,
	urlcolor=blue
}

\begin{document}
\title{Chiral currents in Bose-Einstein condensates subject to current-density interactions}
	
\author{Maria Arazo}
\affiliation{
	Departament de F\'isica Qu\`antica i Astrof\'isica, Universitat de Barcelona, Mart\'i Franqu\`es 1, 08028 Barcelona, Spain
}
\author{Vicente Delgado}
\affiliation{Departamento de F\'isica, Universidad de La Laguna, Tenerife 38200, Spain}
\author{Montserrat Guilleumas}
\affiliation{
	Departament de F\'isica Qu\`antica i Astrof\'isica, Universitat de Barcelona, Mart\'i Franqu\`es 1, 08028 Barcelona, Spain
}
\author{Ricardo Mayol}
\affiliation{
	Departament de F\'isica Qu\`antica i Astrof\'isica, Universitat de Barcelona, Mart\'i Franqu\`es 1, 08028 Barcelona, Spain
}
\author{Antonio Mu\~{n}oz Mateo}
\affiliation{Departamento de F\'isica, Universidad de La Laguna, Tenerife 38200, Spain}


\begin{abstract} 
Persistent currents in quasi-one-dimensional Bose-Einstein condensates become chiral in the presence of current-density interactions. This phenomenon is explored in ultracold atoms loaded in a rotating ring geometry, where diverse current-carrying stationary states are analytically found to generalize previously known solutions to the mean-field equations of motion. 
Their dynamical stability is tested by numerical simulations that show stable currents for states with both constant and modulated density profiles, while decaying currents appear only beyond a unidirectional velocity threshold.
Recent experiments in the field make these states within experimental reach. 
\end{abstract}

\maketitle

\section{Introduction}

More than two decades ago, a new chiral theory was predicted to emerge from the presence of a density-dependent gauge field~\cite{Aglietti1996}. By performing a unitary transformation, the theory can be mapped into a one-dimensional (1D) Gross-Pitaevskii equation in which the number density, entering the interaction term, is replaced by the current density \cite{Jackiw1997}. Recently, this theory has been experimentally realized in ultracold atoms~\cite{Clark2018,Gorg2019,Yao2022,Frolian2022} with light-induced density-dependent gauge potentials~\cite{dalibard2011,goldman2014,chisholm2022,Edmonds2013}. The chiral properties can be observed in current-carrying states~\cite{Edmonds2013,Edmonds2015,Jia2023}, and are particularly manifest in chiral solitons~\cite{Aglietti1996,griguolo1998,Dingwall2018,Bhat2021,Gao2022,Jia2022}. Quasi-periodic time dynamics, in the presence of both contact and current-density interactions, has also been reported for a chiral bright soliton rotating in a ring~\cite{ohberg2019quantum}.
	
In physics, chirality is usually associated with the combination of spin and motion, and it is particularly relevant in relativistic theories concerned with weak interactions and neutrinos~\cite{Neutrinos}, and also in condensed-matter, non-relativistic systems subject to spin-orbit interactions~\cite{Spintronics}. In addition, there exist spinless systems that show chiral dynamics due to the broken time-reversal symmetry, as the unidirectional motion of topological edge states in the quantum Hall effect~\cite{Halperin1982} and in photonic crystals~\cite{Wang2009}. 

The Gross-Pitaevskii equation with current-density interaction belongs to the latter-mentioned set: it describes a spinless system that is not symmetric under time reversal. In homogeneous settings, plane waves and particular soliton states have been shown to solve this equation of motion \cite{Aglietti1996,Jackiw1997,Edmonds2013}. However, neither their interconnection nor the existence of more generic stationary states has yet been explored. In the present work, we report on general solutions to the equations of motion in a ring configuration. In this regard, our goal is to generalize what has been done with contact interactions~\cite{Kanamoto2009} to systems with current-density interactions. There, the symmetry with respect to Galilean transformations results in a symmetric dispersion of plane waves and solitonic states. Here, though, the Galilean symmetry is missing, thus clockwise or counterclockwise rotations are not equivalent motions, and increasing speeds translate into increasing interactions. As a result, expected asymmetries but also unexpected paths in the dispersion relations arise. Apparent differences between current-density and contact interactions are shown in the trajectories and also in the profile of stationary states. For varying rotation, solitonic states connect, in general, with plane waves, although for negative velocities, after a velocity (or alternatively, an interaction) threshold, some branches of bright solitons are detached and follow a free-particle dispersion. The stability of stationary states is analyzed both by linearization and nonlinear time evolution of the Gross-Pitaevskii equation with current-density interactions.

The rest of the paper is structured in three sections. Section~\ref{sec:Model} goes over the equations of motion and conserved quantities that describe the system dynamics. It also sets the stage around plane waves and asymptotic soliton states. Section~\ref{sec:General} determines the general solutions to the equations of motion, the key parameters for their existence, and their stability. Finally, Sec.~\ref{sec:Conclusion} presents our conclusions.

\section{Theoretical model}
\label{sec:Model}

 
We consider a quasi-one-dimensional, rotating ring of radius $R$, where
the condensate wave function $\psi(x,t)$ follows the generalized Gross-Pitaevskii (GP) equation~\cite{Jackiw1997}
\begin{equation}
i\hbar{\partial_t\psi}=
\frac{1}{2M}(-i\hbar\partial_x-M\Omega R)^2\psi+\kappa\hbar J\,  \psi,
\label{eq:GP}
\end{equation}
where $\Omega$ is the angular rotation rate, $J(x,t)$ is the current density measured in the laboratory frame $J=\hbar(\psi^*\partial_x\psi-\psi\partial_x\psi^*)/(i2M)$, and $\kappa$ (dimensionless) is the strength of the current-dependent interaction.
The wave function is normalized to the number of particles $N$ in the condensate, $N=\oint dx\,|\psi|^2$. 
For later use, we introduce the average number density $n_0=N/(2\pi R)$, and the wave number $k_{\text{\tiny $\Omega$}}=M|\Omega| R/\hbar$ associated with rotation; we will also make use of the non-dimensional quantities $\tilde n=R\,n_0$, for the density, and $\tilde \Omega=M R^2\Omega/\hbar$, for the angular rotation.


The stationary states $\psi(x,t)=\phi(x)\,\exp(-i\mu t/\hbar)$ fulfill the time-independent equation
\begin{equation}
\hat H\; \phi = \mu\, \phi,
\label{eq:GPe}
\end{equation}
where $\mu$ is the energy eigenvalue of the nonlinear Hamiltonian operator $\hat H=(-i\hbar\partial_x-M\Omega R)^2/2M+\kappa\hbar J$. 
However, the conserved energy $E$ is given by the expectation value of just the first term in $\hat H$, that is~\cite{Aglietti1996,Jia2022}
\begin{equation}
E = \frac{1}{2M} \oint dx\,\psi^*(-i\hbar\partial_x-M\Omega R)^2\psi.
\label{eq:E}
\end{equation}


Since the system considered is translational invariant, Eq.~\eqref{eq:GP} admits plane-wave solutions
\begin{equation}
\psi_q(x,t)=\sqrt{n_0}\,e^{i(q\,x-\mu_q t/\hbar)},
\label{eq:pw}
\end{equation}
with wave number $q$, such that $qR=0,\pm 1,\pm2,\dots$, and energy eigenvalue
\begin{equation}
\mu_q=\frac{(\hbar q-M\Omega R)^2}{2M}+\kappa\frac{\hbar^2 q \,n_0}{M}.
\label{eq:muPW}
\end{equation} 
As can be seen, the energy shift $\kappa\hbar^2 q \,n_0/M$ of the plane-wave eigenvalue with respect to the non-interacting system ($\kappa=0$) increases in absolute value with the number density. 


Other than plane waves, dark ($\psi_\text{\tiny D}$) and bright ($\psi_\text{\tiny B}$) soliton solutions can be found in the literature~\cite{Aglietti1996}, which, for completeness, we rewrite here as approximate stationary states (when their characteristic lengths $\xi_\text{\tiny D}$ and $\xi_\text{\tiny B}$ are small against the radius) in the ring moving with angular velocity $\Omega$:
\begin{equation}
  \psi_\text{\tiny D}(x,t) \approx \sqrt{\frac{N}{2(\pi R-\xi_\text{\tiny D})}} \; \tanh\left({x}/{\xi_\text{\tiny D}}\right)\, e^{i\left(k_\text{\tiny$\Omega$} x-\mu_\text{\tiny D} t/\hbar\right)},
\label{eq:Dsol}
\end{equation}
if $\Omega>0$, and 
\begin{equation}
  \psi_\text{\tiny B}(x,t) \approx \sqrt{\frac{N}{2\xi_\text{\tiny B}}} \; \mathrm{sech}\left({x}/{\xi_\text{\tiny B}}\right)\, e^{-i\left(k_\text{\tiny$\Omega$} x+\mu_\text{\tiny B} t/\hbar\right)},
\label{eq:Bsol} 
\end{equation}
if $\Omega<0$. 
The characteristic lengths are $\xi_\text{\tiny B}=2( \kappa  N k_{\text{\tiny $\Omega$}})^{-1}$, and $\xi_\text{\tiny D}=\xi_\text{\tiny B}\,(\sqrt{1+4\pi R/\xi_\text{\tiny B}}-1)/2$, and the energy eigenvalues are
\begin{eqnarray}
  \mu_\text{\tiny D} &\approx& \frac{\hbar^2}{M\xi_\text{\tiny D}^2}=\frac{\kappa N\, R}{2(\pi R-\xi_\text{\tiny D})}\,\hbar\Omega, \label{eq:muDsol} \\
  \mu_\text{\tiny B} &\approx& \frac{-\hbar^2}{2M\xi_\text{\tiny B}^2}=-\left(\frac{\kappa N}{2}\right)^2\frac{M\,\Omega^2 R^2}{2}. \label{eq:muBsol}
\end{eqnarray}
These expressions show different scaling: the dark soliton eigenvalue (for $\xi_\text{\tiny D}\ll\pi R$) varies linearly with $\kappa \Omega R n_0$, while the bright soliton eigenvalue (for $\xi_\text{\tiny B}\ll \pi R$) scales quadratically in $\kappa \Omega n_0$ and with the fourth power of the ring radius $R$. 
Another difference between $\psi_\text{\tiny D}$ and $\psi_\text{\tiny B}$ comes from their domains of existence.
In the regime considered, the bright soliton exists for arbitrary values of $|\Omega|$; in this respect, it resembles the dynamics of a classical particle. 
However, the dark soliton presents a particular constraint in the ring since its $\pi$-phase jump has to be canceled by a background constant velocity $\Omega R =\hbar (\pi+2\pi j)/(2\pi R M)$, with $j=0,\pm 1,\pm 2,...$, which restricts the possible rotation rates to $\Omega_\text{\tiny D}=(1/2+j)\hbar/MR^2$.  This boundary condition remarks the wave character of the dark soliton.
 
In what follows, for a given average density $n_0$ (or for a given particle number $N$), we search for analytical stationary solutions to Eq.~\eqref{eq:GPe} that generalize the known solutions, Eqs.~\eqref{eq:pw}, \eqref{eq:Dsol} and~\eqref{eq:Bsol}, where the two latter ones correspond to the asymptotic limit for large rings.

\section{General current-carrying states}
\label{sec:General}

\begin{figure}[tb]
	\flushleft ({\bf a})\\ 
	\includegraphics[width=\linewidth]{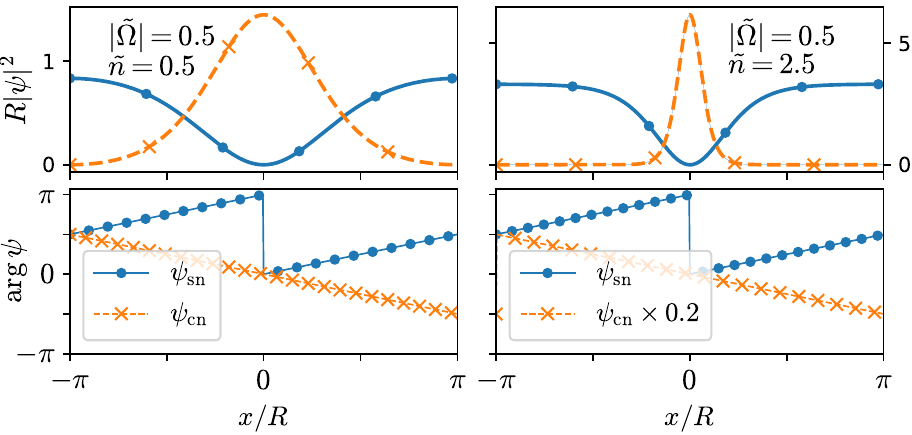}\\ \vspace{-0.2cm}
	({\bf b})\\ 
	\includegraphics[width=\linewidth]{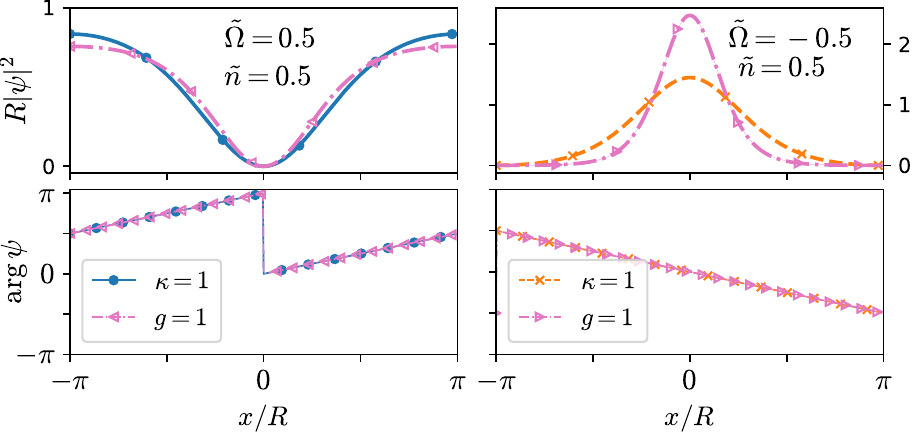}
	\caption{Stationary solitonic states defined by Eqs.~\eqref{eq:Dsn} and~\eqref{eq:Bcn} for $|\Omega|=0.5\,\hbar/(M R^2)$. (a)  States at low  $n_0=0.5/R$ (left panels) versus high  $n_0=2.5/R$ (right panels) number density; they correspond to the black solid symbols in Fig.~\ref{fig:planewaves}. (b) For $n_0=0.5/R$, comparison with equivalent states obtained with contact interaction (of non-dimensional, nonlinear strentgh $g$) for the same interaction strength $g=\kappa=1$; dark soliton with $\Omega=0.5\,\hbar/(M R^2)$ (left panel) and bright soliton with $\Omega=-0.5\,\hbar/(M R^2)$ (right panel).
	The labels indicate the dimensionless values of $\Omega$ and $n_0$, with units $\hbar/(M R^2)$ and $1/R$, respectively.}
	\label{fig:sn_cn_n05}
\end{figure}

The approximate solutions for large $R$, Eqs.~\eqref{eq:Dsol} and~\eqref{eq:Bsol}, can be made exact for generic rings by means of Jacobi elliptic functions, which are periodic functions that generalize the trigonometric functions~\cite{Abramowitz}. 
In particular, the Jacobi sine function ${\rm sn}(x/\xi,\,\mathfrak{m})$, with characteristic width $\xi$ and parameter $\mathfrak{m}\in[0,\,1]$, generalizes the dark soliton solution~\eqref{eq:Dsol} if $\Omega>0$, 
\begin{equation}
  \psi_{\rm sn}(x,t)=\sqrt{\frac{\mathfrak{m}}{\kappa\, \xi^2 k_{\text{\tiny $\Omega$}}}} \; {\rm sn}\left({x}/{\xi},\,\mathfrak{m}\right)\, e^{i(  k_{\text{\tiny $\Omega$}} x-\mu_{\rm sn} t/\hbar)},
  \label{eq:Dsn}
\end{equation}
while the Jacobi cosine function ${\rm cn}(x/\xi,\,\mathfrak{m})$ generalizes the bright soliton solution~\eqref{eq:Bsol} if $\Omega<0$,
\begin{equation}
  \psi_{\rm cn}(x,t)=\sqrt{\frac{\mathfrak{m}}{\kappa\, \xi^2 k_{\text{\tiny $\Omega$}}}} \; {\rm cn}\left({x}/{\xi},\,\mathfrak{m}\right)\, e^{-i(k_{\text{\tiny $\Omega$}} x+\mu_{\rm cn} t/\hbar)},
\label{eq:Bcn} 
\end{equation}
and the energy eigenvalues are
\begin{eqnarray}
  \mu_{\rm sn} &=& \frac{\hbar^2}{2M \xi^2}(1+\mathfrak{m}), \label{eq:muDsn} \\
  \mu_{\rm cn} &=& -\frac{\hbar^2}{2M \xi^2}(2\mathfrak{m}-1). \label{eq:muBcn}
\end{eqnarray}
In both cases, since the phase is a linear function of the position and follows the ring motion, the current density vanishes in the co-moving reference frame. 
An example is shown in Fig.~\ref{fig:sn_cn_n05}(a), which depicts the states given by Eqs.~\eqref{eq:Dsn} and~\eqref{eq:Bcn} for a single node (one soliton) along the ring, rotation rate $|\Omega|=0.5\,\hbar/(M R^2)$, and two values of the average number density, $n_0=0.5/R$ (left panels) and $n_0=2.5/R$ (right panels). 
As can be seen, increasing the number of particles, thus the interaction, translates into narrower solitons. 
For given average density and rotation, these states show clear differences [see Fig.~\ref{fig:sn_cn_n05}(b)] with the corresponding states in the regular Gross-Pitaevskii theory with contact interactions.  

For $\mathfrak{m}\rightarrow 1$, the elliptic wave functions~\eqref{eq:Dsn} and~\eqref{eq:Bcn} tend to the hyperbolic functions~\eqref{eq:Dsol} and~\eqref{eq:Bsol}, respectively. 
In the opposite limit, for $\mathfrak{m}\rightarrow 0$, the trigonometric functions $\sin(x/\xi)$ and $\cos(x/\xi)$ are obtained.
The spatial period of the elliptic functions is given in terms of the complete elliptic integral of the first kind $K(\mathfrak{m})$~\cite{Abramowitz}. Since the period has to be a divisor of the ring length $2\pi R$, it fulfills
\begin{equation}
j\,K(\mathfrak{m})\,\xi={\pi R},
\label{eq:period}
\end{equation}
where $j=1,2,3,\dots$ is a positive integer that indicates the number of nodes in the ring.
Equation~\eqref{eq:period} is also the definition of the characteristic length of the corresponding nonlinear wave. 
Imposing the phase periodicity leads to the possible rotation rates where this solution can exist: $|\Omega|=j\,\hbar/(2MR^2)$. 
By means of Eq.~\eqref{eq:period}, the wave-function normalization provides an implicit equation for the parameter $\mathfrak{m}$:
\begin{eqnarray}
  k_{\text{\tiny $\Omega$}}\,\frac{N_{\rm sn}}{2\pi R} &=& \frac{L(\mathfrak{m})}{\kappa\,\xi^2},
\label{eq:Dnorm}\\
  k_{\text{\tiny $\Omega$}}\,\frac{N_{\rm cn}}{2\pi R} &=& \frac{\mathfrak{m}-L(\mathfrak{m})}{\kappa\,\xi^2 },
\label{eq:Bnorm}
\end{eqnarray}
where we have defined the function $L(\mathfrak{m})=1-E(\mathfrak{m})/K(\mathfrak{m})$, and $E(\mathfrak{m})$ is the complete elliptic integral of the second kind. 
$L(\mathfrak{m})$ takes values in the range $[0,1]$ when $\mathfrak{m}$ varies in $[0,1]$.

\subsection{General solutions}

More general solutions to Eq.~\eqref{eq:GP} can be found that interpolate between plane waves~\eqref{eq:pw} and soliton-like states~\eqref{eq:Dsn} and~\eqref{eq:Bcn}.
Generic stationary states $\psi(x,t)=\sqrt{n(x)}\,\exp[i\theta(x)-i\mu t/\hbar]$ have number density $n(x)$ and phase $\theta(x)$ given in terms of the system parameters $\{R,N,\Omega,\kappa\}$.
The condensate phase is related to the density through the (stationary) continuity equation in the moving reference frame $\partial_x (J-n\Omega R)=0$, which gives rise to the constant current density
\begin{equation}
  \mathcal{J}=n\left(\frac{\hbar}{M}\partial_x\theta-\Omega R\right).
\label{eq:J}
\end{equation} 
Notice that $\mathcal{J}=0$ for both the plane waves~\eqref{eq:pw} and the soliton-like states defined in Eqs.~\eqref{eq:Dsn} and~\eqref{eq:Bcn}.
By using Eq.~\eqref{eq:J}, the GP Eq.~\eqref{eq:GPe} can be written as a single equation for the density, 
\begin{equation}
  \mu=-\frac{\hbar^2}{2M}\frac{\partial_x^2 \sqrt{n}}{\sqrt{n}}+\frac{M}{2}\left(\frac{\mathcal{J}}{n}\right)^2+\kappa\hbar(\mathcal{J}+\Omega R n).
\label{eq:GPdensity}
\end{equation}
Premultiplication by $\partial_x \sqrt{n}$ and subsequent integration brings the density equation to the form
\begin{equation}
  \left(\partial_x n\right)^2 \hspace{-0.1em} = \hspace{-0.1em} \frac{4M}{\hbar^2} 
  \left[ \hbar \kappa \Omega R n^3 \hspace{-0.1em} - \hspace{-0.1em} 
  2(\mu - \hbar\kappa\mathcal{J})n^2 \hspace{-0.1em} - \hspace{-0.1em} 
  2 C n \hspace{-0.1em} - \hspace{-0.1em} 
  M \mathcal{J}^2 \right], 
\label{eq:GPdensity_general}
\end{equation}
where $C$ is an integration constant with units of energy per unit length. 
Since the right-hand side of Eq.~\eqref{eq:GPdensity_general} is a cubic polynomial in the density, the formal, general solution of this equation is the Weierstrass $\wp$ function~\cite{Abramowitz} (see Appendix). 
In particular cases, the Weierstrass function can be written as Jacobi elliptic functions, which arise from the transformation $n(x)=\alpha + \beta f^2(x)$ and the constraint
\begin{equation}
M\mathcal{J}^2-{2\kappa\hbar \alpha^2}\mathcal{J}-\left(\kappa\hbar\Omega R \alpha^3-2\mu
\alpha^2-2C\alpha\right)=0.
\label{eq:dnJ0}
\end{equation}

At this stage, following Refs.~\cite{carr2000a,carr2000b,Kanamoto2009}, we choose $f(x)={\rm dn}(x/\xi,\mathfrak{m})$, the Jacobi ${\rm dn}$ function with parameter $\mathfrak{m}$ and argument $x/\xi$, so that (see Appendix for details)
\begin{equation}
  \psi(x,t)=\sqrt{\alpha+\beta\,{\rm dn}^2(x/\xi,\mathfrak{m})}\;e^{i\theta(x)-i\mu t/\hbar},
\label{eq:general}
\end{equation}
where the characteristic width $\xi$ satisfies
\begin{equation}
  \xi=\frac{\hbar}{\sqrt{ M \kappa\hbar|\Omega \beta| R}}.
\label{eq:dnWidth}
\end{equation}
The phase, from the integration of Eq.~\eqref{eq:J},
\begin{equation}
  \theta(x)={{\rm sgn}(\Omega)}\,{k_{\text{\tiny $\Omega$}}}{x} +\frac{M\xi\mathcal{J}}{\hbar(\alpha+\beta)}\Pi(\eta;x/\xi,\mathfrak{m}),
\label{eq:dnPhase}
\end{equation}
is expressed in terms of the incomplete elliptic integral of the third kind $\Pi(\eta;x/\xi,\mathfrak{m})$~\cite{Abramowitz}, with $\eta=\mathfrak{m}\beta/(\alpha+\beta)$.
For $\mathfrak{m}\rightarrow 0$, since ${\rm dn}(x/\xi,\mathfrak{m})\rightarrow 1$, the wave function~\eqref{eq:general} approaches the plane-wave solutions~\eqref{eq:pw}, whereas for $\mathfrak{m}\rightarrow 1$, ${\rm dn}(x/\xi,\mathfrak{m})\rightarrow {\rm sech}(x/\xi)$ and the wave function can approach the soliton solutions~\eqref{eq:Dsol} and~\eqref{eq:Bsol} depending on the values of $\alpha$ and $\beta$. 
In particular, Eq.~\eqref{eq:general} gives dark-soliton states ($\Omega>0$) for $\beta<0$ and bright-soliton states ($\Omega<0$) for $\beta>0$, so $\Omega\beta<0$ for all cases.

The expression for the wave function period, Eq.~\eqref{eq:period}, determines $\beta$ (implicit in $\xi$), and then $\alpha$ can be calculated from the state normalization. 
From them, one obtains the energy eigenvalue
\begin{equation}
  \mu=\left(\mathfrak{m}-2-3\frac{\alpha}{\beta}\right)\frac{\hbar^2}{2M\xi^2}+\kappa\hbar\mathcal{J},
\label{eq:dn_mu}
\end{equation}
which resembles (and reduces in the proper limits to) Eqs.~\eqref{eq:muDsn} and~\eqref{eq:muBcn}, and the constant current density
\begin{equation}
\mathcal{J}=\pm\frac{\hbar\beta }{M\xi}\sqrt{
	(\mathfrak{m}-1)\frac{\alpha}{\beta}+(\mathfrak{m}-2) \frac{\alpha^2}{\beta^2}-\frac{\alpha^3}{\beta^3}}.
\label{eq:dnJ1}
\end{equation} 
On the other hand, the periodicity of the phase, by substitution of the expressions for $\xi$, $\alpha$, $\beta$, $\mu$, and $\mathcal{J}$, results in an implicit equation for the parameter $\mathfrak{m}$ in terms of the system quantities $\{R,N,\Omega,\kappa\}$ (see Appendix).

The general solution, Eq.~\eqref{eq:general}, leads to soliton states only if $\Omega\neq 0$. 
When the rotation rate approaches zero, the length scale $\xi$ associated with the soliton grows infinitely, and so Eq.~\eqref{eq:general} only gives plane-wave solutions. 
Still, the Weierstrass $\wp$ function is the formal solution as $\Omega\rightarrow 0$, but it diverges in this case, and such a solution with infinite density has no physical meaning. 
In any case, the non-rotating scenario is actually a singular case of Eq.~\eqref{eq:GPdensity_general} since the cubic term of the equation vanishes for $\Omega=0$, the equation reduces to the Schr\"odinger equation, and the Weierstrass and Jacobi elliptic functions are no longer its solution.

\subsection{Case study}
\label{sec:case}

\begin{figure}[tb]
	\includegraphics[width=\linewidth]{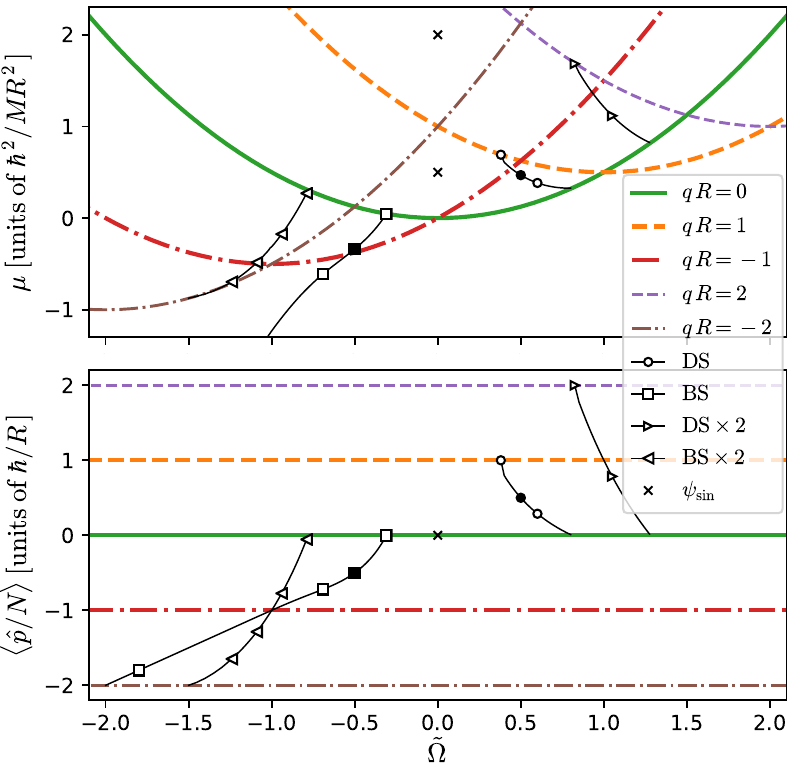}
	\caption{Energy eigenvalues $\mu$ (top panel) and average momentum per particle (bottom panel) of plane-wave states (thick lines) with wave numbers $q\in[-2,\,2]\times 1/R$, and dark (DS) and bright (BS) soliton-like states (thin lines with symbols) in a rotating ring with number density $n_0=0.5/R$ and current-density-interaction strength $\kappa=1$. Isolated states $\psi_{\rm sin}\propto \sin(qx)$ (crosses) are also shown at $\Omega=0$. The states labeled as $\mathrm{DS}\times 2$ ($\mathrm{BS}\times 2$) have two dark (bright) solitons.}
	\label{fig:planewaves}
\end{figure}

\begin{figure}[tb]
	\flushleft 	({\bf a})\\
	\includegraphics[width=\linewidth]{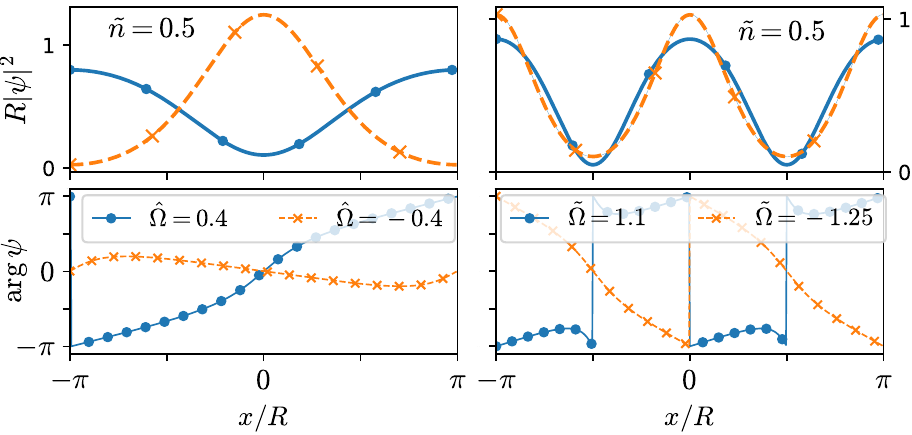}\\
	({\bf b})
	\includegraphics[width=\linewidth]{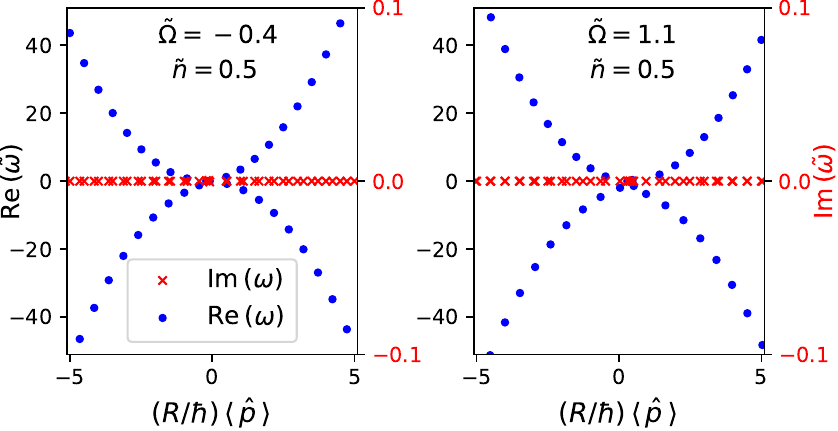}
	\caption{General solitonic states~\eqref{eq:general}, which have a non-zero current $\mathcal{J}\neq 0$ in the rotating frame. (a) States with one soliton (left) and two solitons (right) at a number density $n_0=0.5/R$ for positive and negative rotating rates. (b) Spectrum of linear excitations for a one-bright-soliton state (left) and a two-dark-soliton state (right) with frequency $\omega$ in units of $\hbar/(M R^2)$. The dispersion relations only present real frequencies, indicating that both cases are linearly stable.}
	\label{fig:dn}
\end{figure}

To analyze the general solutions, we focus on a particular case with number density $n_0=0.5/R$ and current-density interaction strength $\kappa=1$. 
Figure~\ref{fig:planewaves} collects the energy eigenvalue $\mu$ and the average (canonical) momentum per particle $\langle{\hat p/N}\rangle=M\oint dx\, J/N$, as a function of the angular velocity $\Omega$, of plane-wave (thick lines) and soliton states (thin lines with symbols); dark and bright solitons belong to trajectories with positive and negative angular velocities, respectively.
All the trajectories show bifurcation points that interconnect families of plane-wave and soliton states;  generally, in contrast to systems with contact interactions, these connections  are not smooth (giving rise to tangent trajectories), since neither $\mu$ nor $\langle{\hat p}\rangle$ retain here their usual meaning of chemical potential and conserved momentum, respectively. On the contrary, the connections of trajectories are smooth in the energy versus rotation graph (see Appendix).

The plane-wave solutions trace parabolas centered at $\Omega/(\hbar/MR^2)=q R$ that have an energy shift of $\kappa \hbar^2 q {n_0}/M$ due to the current-density interactions. 
This asymmetry stems from the chirality of the system and shows as well in the soliton trajectories.
Among the latter, families of states with one or two dark solitons make the connection between two plane-wave trajectories whose non-dimensional wave numbers, $qR$, differ in one or two units, respectively; thus, for instance, the one-dark-soliton family (thin line with circles) connects the plane waves that have $q R =0$ with those with $q R = 1$ [the filled circle corresponds to the particular solution~\eqref{eq:Dsn} that only exists for $\Omega=0.5\,\hbar/(MR^2)$], while the two-dark-soliton family (thin lines with right-pointing triangles) connects $q R =0$ and $q R = 2$. 
In general, as one may expect, solutions with $j$ solitons connect plane-wave states with a difference of $j$ in the non-dimensional wave number $qR$. 

A similar connecting role could be expected to be played by bright solitons at negative angular velocities. 
We show in Fig.~\ref{fig:planewaves} the trajectories for one-bright-soliton states (thin line with squares) and for two bright solitons (thin line with left-pointing triangles). 
As can be seen, the two-soliton family indeed connects the plane-wave solutions with $q R =0$ and $q R = -2$.
The single-soliton trajectory, however, behaves differently: instead of connecting the cases with $q R =0$ and $q R = -1$, it detaches from the expected path as $|\Omega|$ increases (i.e., as the interaction becomes more attractive); the filled square corresponds to the particular solution~\eqref{eq:Bcn} for $\Omega=-0.5\,\hbar/(MR^2)$. 
For high velocities, the single-soliton trajectory reproduces the behavior of a free particle, with energy and momentum varying quadratically and linearly, respectively, with the rotation rate. 
This does not happen for the two-soliton trajectory in the present case since the total attractive interaction (or number of particles) is not large enough for each soliton to reach the free particle features, but this may eventually occur with the right choice of parameters.
The threshold velocity (and interaction) above which bright-soliton states show a free-particle dispersion, therefore, will mainly depend on the number of solitons and on the particular parameters of the system (the number density and the strength of the current-density interaction).

Typical details of the general solitonic states are shown in Fig.~\ref{fig:dn}(a); the density and phase profiles of one-soliton-like (left panels) and two-soliton-like states (right panels) are represented for selected positive and negative values of the rotation rate. 
By contrast with the solutions in Fig.~\ref{fig:sn_cn_n05}, these states present a constant, non-vanishing current $\mathcal{J}\neq 0$ in the rotating frame. 
As can be inferred from the bottom panel of Fig.~\ref{fig:planewaves}, and by means of Eq.~\eqref{eq:J} so that $2\pi R\,\mathcal{J}=\langle \hat p/M\rangle-\Omega R\,N$, for a family with $j$ solitons, $\mathcal{J}$ vanishes for the particular solutions~\eqref{eq:Bcn} and~\eqref{eq:Dsn}, and also once a bright soliton family reaches the free-particle dispersion, as happens to the one-bright-soliton family in the present case; moving away from these particular cases, $\mathcal{J}$ decreases (increases) for increasing (decreasing) rotation rate  when $\Omega$ is positive (negative). This fact reflects the production of backflow currents in response to the soliton phase jumps.

In the non-rotating case, $\Omega=0$, one can find plane-wave states along with the isolated, sinusoidal solutions, $\psi_{\rm sin}\propto \sin(q\,x)$ and $\psi_{\rm cos}\propto \cos(q\,x)$, marked as crosses in Fig.~\ref{fig:planewaves}, which (due to the absence of current) are degenerate solutions of the Schr\"odinger equation; thus, superpositions of these states with real amplitudes are also allowed solutions.
As we mentioned earlier, the general solution that we used so far, Eq.~\eqref{eq:general}, only yields plane-wave states at this point, and the more general solution, the Weierstrass $\wp$ function, is not physically valid. 
This means that one cannot follow a connection path between plane-wave families with wave numbers of different signs, $q R > 0$ and $q R <0$, as doing so would provide a means of adiabatically changing the chirality of the states. 
Contrary to the situation in systems with contact interactions, there is no such path here, and the sinusoidal solutions that should be part of it, at $\Omega=0$, remain unconnected as isolated states.

\begin{figure}[tb]
	\flushleft 	({\bf a})\\
	\includegraphics[width=\linewidth]{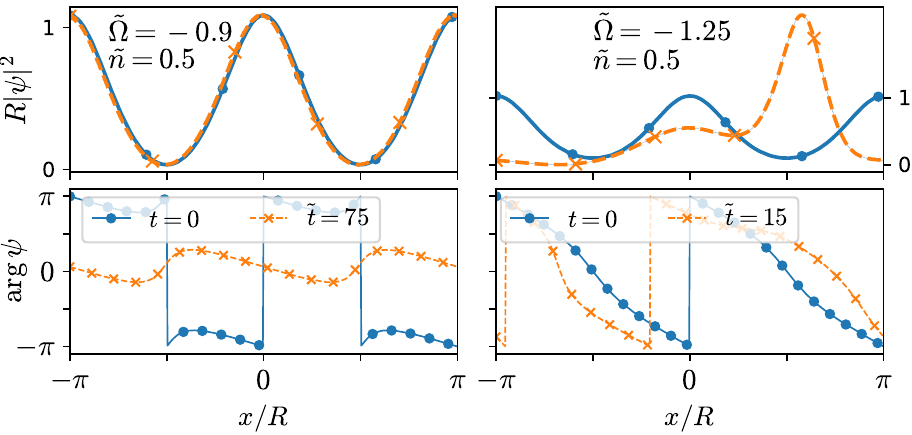}\ 
	({\bf b})
	\includegraphics[width=\linewidth]{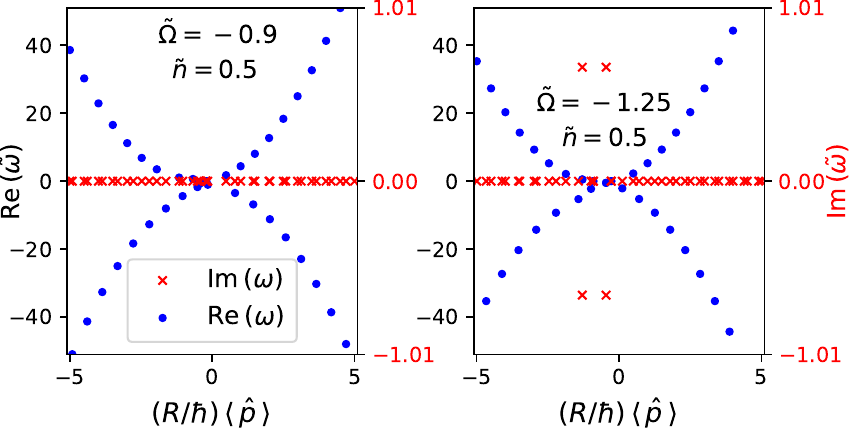}
	\caption{Stable (left panels) and unstable (right panels) states with two bright solitons. (a) Initial stationary states and final states after a nonlinear evolution; see Fig.~\ref{fig:evol_real} for more details. (b) Spectra of linear excitations. The state with rotating rate $|\tilde\Omega|<1.0$ is unstable.}	
	\label{fig:evol}
\end{figure}

\begin{figure}[tb]
	\flushleft 	({\bf a})\\
	\includegraphics[width=0.98\linewidth]{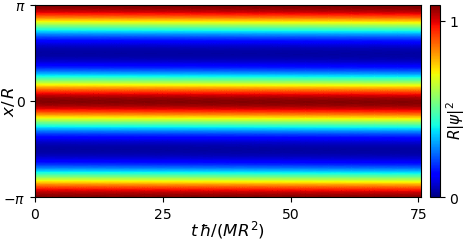}\ 
	({\bf b})
	\includegraphics[width=0.98\linewidth]{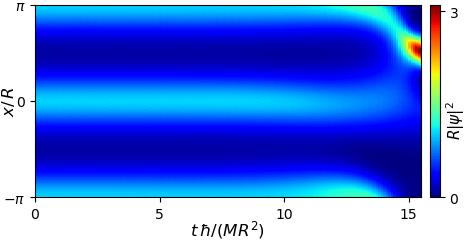}
	\caption{Nonlinear time evolution of the two-bright-soliton states from Fig.~\ref{fig:evol}, with (a) $\tilde\Omega=-0.9$, which is stable, and with (b) $\tilde\Omega=-1.25$, which is unstable. The evolution is done after adding a perturbative amount of white noise over the initial stationary state that changes the energy a 2\%.} 
	\label{fig:evol_real}
\end{figure}

\subsection{Dynamical stability}

To determine the dynamical stability of the states considered, we have performed both the analysis of linear excitation modes and the numerical simulation of the nonlinear time evolution with Eq.~\eqref{eq:GP}. 
To carry out the former analysis, we have solved numerically the Bogoliubov equations for the linear excitations of solitonic states and searched for complex excitation frequencies that could point to dynamical instabilities. 


The linear excitation modes $\delta \psi_j= [u_j,\,v_j]^T$ of a stationary state $\psi$ can be calculated as solutions to the Bogoliubov equations, which are obtained from the substitution in Eq.~\eqref{eq:GP} of the perturbed state $\psi(x,t)= \exp(-i\mu t/\hbar)\,\{\phi(x)+\sum_j [u_j(x)\,\exp(-i\omega_j t)+v_j(x)^*\,\exp(i\omega_j^* t)]\}$, where $j$ indexes the modes. The resulting system of equations is 
\begin{equation}
\hat B\,\delta\psi_j=\hbar\omega_j\,\delta\psi_j,
\label{eq:Bog}
\end{equation} 
with the Bogoliubov operator
\begin{eqnarray}
\hat B &=& \begin{pmatrix}\hat H-\mu &0\\0&-\hat H^*+\mu \end{pmatrix} \nonumber\\
&+& \frac{\hbar\kappa}{2M}
\begin{pmatrix}
\phi\,\left[\phi^\ast\hat{p}-(\hat{p}\,\phi^\ast)\right] &
-\phi\,\left[\phi\,\hat{p}-(\hat{p}\,\phi)\right] \\
-\phi^\ast\left[\phi^\ast\hat{p}-(\hat{p}\,\phi^\ast)\right] &
\phi^\ast\left[\phi\,\hat{p}-(\hat{p}\,\phi)\right]
\end{pmatrix},\quad
\label{eq:Bog1}
	\end{eqnarray}
where  $\hat H$ is the energy operator in Eq.~\eqref{eq:GPe} and $\hat{p}$ is the momentum operator. The dynamical instabilities correspond to modes with complex frequencies ${\rm Im}(\omega_j)\neq 0$.


On the other hand, the dispersion relation for the linear excitation modes of plane waves with wave number $q$ has the analytical expression
\begin{equation}
\omega_k=\frac{\hbar k}{M} \left[
	q \hspace{-0.1em} + \hspace{-0.1em} 
	\frac{\kappa n_0}{2} \pm 
	\sqrt{q \kappa n_0 \hspace{-0.1em} + \hspace{-0.1em} 
	\frac{(\kappa n_0)^2 \hspace{-0.1em}+\hspace{-0.1em} k^2}{4}} \hspace{-0.1em}- \hspace{-0.1em} 
	{\rm sgn}(\Omega)\,k_{\tiny \Omega} \right],
\end{equation}
where $k$ is the wave number of the excitation. 
While all positive $q$ lead to real frequencies, negative $q$ such that $|q|>\kappa n_0/4$ are unstable. In our case study, this means $|q|R>0.125$; therefore, all the plane waves with negative wavenumber are unstable. 
This result suggests that also the solitonic states connecting plane-wave families with negative wave numbers should contain unstable states, at least near the connection points of the corresponding trajectories in the $\mu$ versus $\Omega$ diagram.


For solitonic states with positive rotation and also for the family of one-bright-soliton states with negative rotation, we have not found unstable modes among the spectrum of linear excitations; two typical examples of the corresponding dispersion are shown in Fig.~\ref{fig:dn}(b), which present only real frequencies. 
However, as expected from the above analysis of plane waves, we have found that for negative rotation with $|\tilde\Omega|\ge 1.0$, the two-bright-soliton states have excitation modes with complex frequencies and therefore are (linearly) unstable; in contrast, those states in the same family with $|\tilde\Omega|< 1.0$ are stable. 
Figures~\ref{fig:evol} and~\ref{fig:evol_real} show, respectively, the results of the linear analysis and the nonlinear time evolution for two of these states.
The latter has been obtained after seeding a perturbative amount of white noise on the initial stationary states. The equilibrium configuration is preserved for a long time when $\tilde\Omega=-0.9$; see Fig.~\ref{fig:evol_real}(a). However, when $\tilde\Omega=-1.25$, the real-time evolution shows the decay of the initial two-soliton state, as can be seen in Fig.~\ref{fig:evol_real}(b).
These results confirm the predictions of the linear analysis.

\section{Conclusions}
\label{sec:Conclusion}

We have studied persistent currents that are chiral in quasi-1D Bose-Einstein condensates with current-density interactions loaded in a  rotating ring trap.
The chiral current-carrying states, plane waves and multi soliton-like states, manifest clear differences with respect to systems with contact interparticle interactions. 
The state eigen-energy versus rotation diagram is asymmetric against the direction of the rotation rate $\Omega$, and,
due to this asymmetry, there are no solitonic trajectories crossing $\Omega=0$, so it is not possible to adiabatically connect plane-wave states with different chirality.
%
%
Our tests of dynamical stability, which show agreement between the linear and the nonlinear analisys, demonstrate
stable currents for (positive-velocity) states with both constant and modulated density profiles, 
while decaying currents appear only beyond a unidirectional (negative) velocity threshold. The latter instability points to alternative stable states, moving and strongly localized bright solitons, whose dynamics resemble the energy and momentum features of classical particles.
These results open prospects of new work in the search of equivalent current-carrying states within the 3D framework of the recently realized effective chiral theory  \cite{Frolian2022}.


\begin{acknowledgments}

We thank Marcel Nicolau for the insightful discussions.
This work has been funded by Grant No.~PID2020-114626GB-I00 from the MICIN/AEI/10.13039/501100011033 and by the European Union Regional Development Fund within the ERDF Operational Program of Catalunya (project QuantumCat, Ref. No. 001-P001644).
M. A. is supported by FPI Grant PRE2018-084091. V. D. acknowledges support from Agencia Estatal de Investigaci\'on (Ministerio de Ciencia e Innovaci\'on, Spain) and Fondo Europeo de Desarrollo Regional (FEDER, EU) under grant PID2019-105225GB-100.

\end{acknowledgments}

\appendix

\section*{Appendix: Solutions as Jacobi elliptic functions} 

With the transformation $n(x)=\alpha + \beta f^2(x)$, the density Eq.~\eqref{eq:GPdensity_general} takes the form
\begin{eqnarray}
\left(\partial_x f\right)^2 &=& 
\frac{M\kappa\Omega R \beta}{\hbar} f^4 
- \frac{2M}{\hbar^2}\left(\mu - \hbar \kappa \mathcal{J} -\frac{3\hbar\kappa\Omega R \alpha}{2} \right) f^2 \nonumber \\
&& - \frac{4 M \alpha}{\hbar^2 \beta}\left( \mu - \hbar\kappa\mathcal{J} + \frac{C}{2\alpha} - \frac{3\hbar\kappa\Omega R \alpha}{4}  \right) ,
\end{eqnarray}
if the condition
\begin{equation}
M\mathcal{J}^2-{2\kappa\hbar \alpha^2}\mathcal{J}-\left(\kappa\hbar\Omega R \alpha^3-2\mu
\alpha^2-2C\alpha\right)=0
\label{eq:dnJ0}
\end{equation}
is fulfilled.  
Thus, by means of the Jacobi elliptic function $f(x)={\rm dn}(x/\xi,\mathfrak{m})$, the generalized Gross-Pitaevskii Eq.~\eqref{eq:GP} is satisfied with parameters
\begin{eqnarray}
  \xi &=& \frac{\hbar}{\sqrt{ M \kappa\hbar|\Omega \beta| R}}, 
      \label{eq:dnWidth}\\
  \mathcal{J} &=& \pm\frac{\hbar\beta }{M\xi}
      \sqrt{(\mathfrak{m}-1)\frac{\alpha}{\beta}+(\mathfrak{m}-2) \frac{\alpha^2}{\beta^2}-\frac{\alpha^3}{\beta^3}},
	\label{eq:dnJ1}\\
  \mu &=& \left(\mathfrak{m}-2-3\frac{\alpha}{\beta}\right)\frac{\hbar^2}{2M\xi^2}+\kappa\hbar\mathcal{J}, 
      \label{eq:dn_mu}\\
  C &=& \frac{\hbar^2\beta}{2M\xi^2}\left[1-\mathfrak{m}+2(2-\mathfrak{m})\frac{\alpha}{\beta}+ 3\frac{\alpha^2}{\beta^2}\right],
      \label{eq:dnC}
\end{eqnarray}
and $\Omega\beta<0$.

The condition for the periodicity of the density, Eq.~\eqref{eq:period}, along with Eq.~\eqref{eq:dnWidth}, gives
\begin{equation}
|\beta|=\left(\frac{j K(\mathfrak{m})}{\pi R}\right)^2\frac{1}{\kappa k_{\text{\tiny $\Omega$}}},
\label{eq:dnB}
\end{equation}
where $j=1,\,2,\,3,\dots$, and the normalization imposes the relation between $\alpha$ and $\beta$
\begin{equation}
\alpha+\left[1-{L(\mathfrak{m})}\right]\,\beta=n_0.
\label{eq:dnNorm}
\end{equation}

\begin{figure}[b]
	\includegraphics[width=\linewidth]{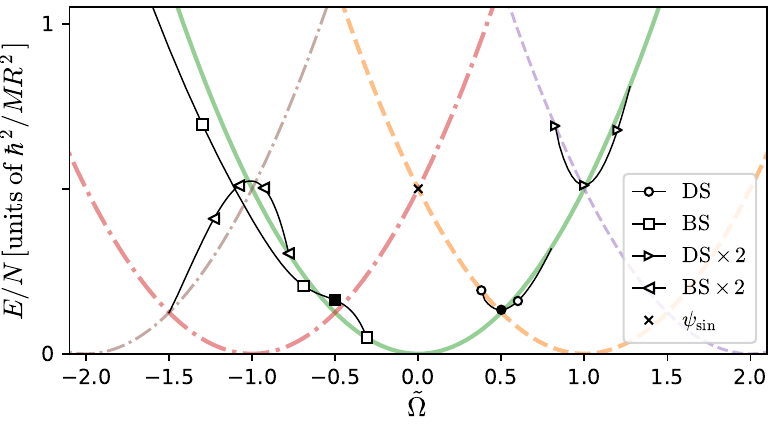}
	\caption{Same as Fig.~\ref{fig:planewaves} but for the energy per particle. The trajectories of solitonic states (thin lines with symbols), as per Eq.~\eqref{eq:dnE}, connect smoothly with the plane-wave trajectories (thick faded lines).}
	\label{fig:energy_n05}
\end{figure}

The phase can be found by means of Eq.~\eqref{eq:J}:
\begin{equation}
\theta(x)={\rm sgn}(\Omega)\,{k_{\text{\tiny $\Omega$}}}{x} +\frac{M\mathcal{J}}{\hbar}\int_0^x  \frac{dx'}{\alpha+\beta{\rm dn}^2(x'/\xi,\mathfrak{m})}.
\label{eq:dnPhase0}
\end{equation}
The latter integral can be expressed in terms of the incomplete elliptic integral of the third kind $\Pi(\eta;x/\xi,\mathfrak{m})$, where $\eta=\mathfrak{m}\beta/(\alpha+\beta)$~\cite{Abramowitz}:
\begin{equation}
\theta(x)={\rm sgn}(\Omega)\,{k_{\text{\tiny $\Omega$}}}{x} +\frac{M\xi\mathcal{J}}{\hbar(\alpha+\beta)}\Pi(\eta;x/\xi,\mathfrak{m}).
\label{eq:dnPhase}
\end{equation}
For $x=2\pi R$, the elliptic integral is complete $\Pi(\eta;\mathfrak{m})$, so that the phase becomes periodic in the ring $\theta(x)=\theta(x+2\pi R)+2\pi l$, with $l$ integer. That is, from Eq.~\eqref{eq:dnPhase}
\begin{equation}
2j\frac{M\xi \mathcal{J}}{\hbar(\alpha+\beta)}\Pi(\eta;\mathfrak{m}) + {\rm sgn}(\Omega) {2\pi R}{k_{\text{\tiny $\Omega$}}}=2\pi l,
\label{eq:dnPhase_period}
\end{equation}
which, by substitution of the expressions for $\xi,\,\alpha,\beta,\mu,\mathcal{J}$, in Eqs.~\eqref{eq:dnWidth}, \eqref{eq:dnNorm}, \eqref{eq:dnB}, \eqref{eq:dn_mu}, and~\eqref{eq:dnJ1}, results in an implicit equation for the parameter $\mathfrak{m}$. Once this is found, substituting backwards, all the constants in Eq.~\eqref{eq:general} are obtained. After substitution in Eq.~\eqref{eq:E}, the resulting energy is
\begin{eqnarray}
\frac{E}{N} &=& \frac{\hbar^2}{2M\xi^2}\left[\mathfrak{m}-2L(\mathfrak{m})-\frac{\alpha}{\beta}-\right.  \nonumber \\
&&  \left. -\frac{2\beta}{3n_0}\left(\mathfrak{m}-2(\mathfrak{m}+1)L(\mathfrak{m})+3L(\mathfrak{m})^2\right)\right] .
\label{eq:dnE}
\end{eqnarray}
Figure~\ref{fig:energy_n05} shows the energies of the states considered in our case study of Sec.~\ref{sec:case}.

\bibliography{chiral_currents}

\end{document}